\documentclass[final,5p,times,twocolumn]{elsarticle}

\usepackage{lineno}
\usepackage[hidelinks]{hyperref}

\usepackage{graphicx}
\usepackage{amsmath}   

\begin{document}

\begin{frontmatter}

\title{ Development of nanocomposite scintillators for use in high-energy physics}

\author[LNF]{A.~Antonelli}
\author[CERN]{E.~Auffray}
\author[Bicocca,GTP]{S.~Brovelli}
\author[Bicocca]{F.~Bruni}
\author[Napoli,NA]{M.~Campajola}
\author[Insubria,MIB]{S.~Carsi\corref{cor1}}
\ead{stefano.carsi.uni@gmail.com}
\author[Bicocca]{F.~Carulli}
\author[Napoli,NA]{G.~De Nardo}
\author[LNF,TorVergata]{E.~Di Meco}
\author[LNF]{E.~Diociaiuti}
\author[Bicocca]{A.~Erroi}
\author[NA]{M.~Francesconi}
\author[CERN,Munich]{I.~Frank}
\author[PI]{S.~Kholodenko}
\author[CERN]{N.~Kratochwil}
\author[RM]{E.~Leonardi}
\author[Insubria,MIB]{G.~Lezzani}
\author[Insubria,MIB]{S.~Mangiacavalli}
\author[LNF]{S.~Martellotti}
\author[NA]{M.~Mirra}
\author[Trieste,TS]{P.~Monti-Guarnieri}
\author[LNF]{M.~Moulson}
\author[LNF,TorVergata]{D.~Paesani}
\author[LNF]{E.~Paoletti}
\author[Insubria,MIB]{L.~Perna}
\author[LNF]{D.~Pierluigi}
\author[Insubria,MIB]{M.~Prest}
\author[Ferrara,FE]{M.~Romagnoni}
\author[LNF]{A.~Russo}
\author[LNF]{I.~Sarra}
\author[Insubria,MIB]{A.~Selmi}
\author[Padova,LNL]{F.~Sgarbossa}
\author[LNF]{M.~Soldani}
\author[LNF]{R.~Tesauro}
\author[LNF]{G.~Tinti}
\author[MIB]{E.~Vallazza}

\address[LNF]{INFN Laboratori Nazionali di Frascati, Frascati, Italy}
\address[CERN]{CERN, Meyrin, Switzerland}
\address[Bicocca]{Università degli Studi di Milano Bicocca, Milan, Italy}
\address[GTP]{Glass to Power SpA, Rovereto, Italy}
\address[Napoli]{Università degli Studi di Napoli Federico II, Naples, Italy}
\address[NA]{INFN Sezione di Napoli, Naples, Italy}
\address[Insubria]{Università degli Studi dell'Insubria, Como, Italy}
\address[MIB]{INFN Sezione di Milano Bicocca, Milan, Italy}
\address[TorVergata]{Università degli Studi di Roma Tor Vergata, Rome, Italy}
\address[Munich]{Ludwig-Maximilians-Universität München, Munich, Germany}
\address[PI]{INFN Sezione di Pisa, Pisa, Italy}
\address[RM]{INFN Sezione di Roma, Rome, Italy}
\address[Trieste]{Università degli Studi di Trieste, Trieste, Italy}
\address[TS]{INFN Sezione di Trieste, Trieste, Italy}
\address[Ferrara]{Università degli Studi di Ferrara, Ferrara, Italy}
\address[FE]{INFN Sezione di Ferrara, Ferrara, Italy}
\address[Padova]{Università degli Studi di Padova, Padua, Italy}
\address[LNL]{INFN Laboratori Nazionali di Legnaro, Legnaro, Italy}

\cortext[cor1]{Corresponding author}

\begin{abstract}
 Semiconductor nanocrystals (“quantum dots”) are light emitters with high quantum yield that are relatively easy to manufacture. There is therefore much interest in their possible application for the development of high-performance scintillators for use in high-energy physics. 
 However, few previous studies have focused on the response of these materials to high-energy particles. To evaluate the potential for the use of nanocomposite scintillators in calorimetry, we are performing side-by-side tests of fine-sampling shashlyk calorimeter prototypes with both conventional and nanocomposite scintillators using electron and minimum-ionizing particle beams, allowing direct comparison of the performance obtained.
\end{abstract}


\end{frontmatter}

Semiconductor nanostructures can be used to produce ultrafast, robust
scintillators. As an example, perovskite nanocrystals have 
been found to have good scintillation light output, to have decay times down to O(100~ps), and to be robust with respect to radiation doses of O(1 MGy)~\cite{Erroi}. These nanocrystals can be cast with a polymer matrix to obtain a nanocomposite (NC) scintillator, which can be engineered to meet the performance requirements for a variety of applications in calorimetry for high-energy physics (see \cite{Doser} and references therein).

Nanocomposite scintillators have received much attention in the materials-science community, with photo- and radio luminescence studies having been performed on samples of many different compositions~\cite{Doser}. However, almost no studies have been done on the response of NC scintillators to high-energy particles, especially single minimum-ionizing particles (mips).

The goal of the NanoCal project~\cite{AIDAinnova} is to construct a calorimeter prototype with NC scintillator and test it with high-energy beams. The shashlyk calorimeter design~\cite{Atoian} is naturally ideal as a test platform: it is easy to construct a shashlyk prototype with very fine sampling, and both the primary scintillator and WLS materials required can be optimized using NC technology.

The design of our first prototypes was inspired by photo- and radioluminescence results obtained with NC scintillators consisting of CsPbBr$_3$ nanocrystals dissolved in a UV-cured polyacrylate matrix at concentrations of 0.05 to 0.8\% w/w \cite{Erroi}. In these studies, the light yield was seen to be linear with concentration, reaching a value of 4800 ph/MeV for the 0.8\% sample. From a 3-component fit to the pulse shape for the 0.2\% sample, more than 30\% of the light was seen to be emitted in $<80$~ps, with about 20\% emitted with a decay time of $\sim$600 ps, and the rest with $\tau\sim10$~ns.
No decrease in the light yield was observed after radiation doses of up to 1~MGy. 

\begin{figure}
\centering
\includegraphics[width=0.25\textwidth]{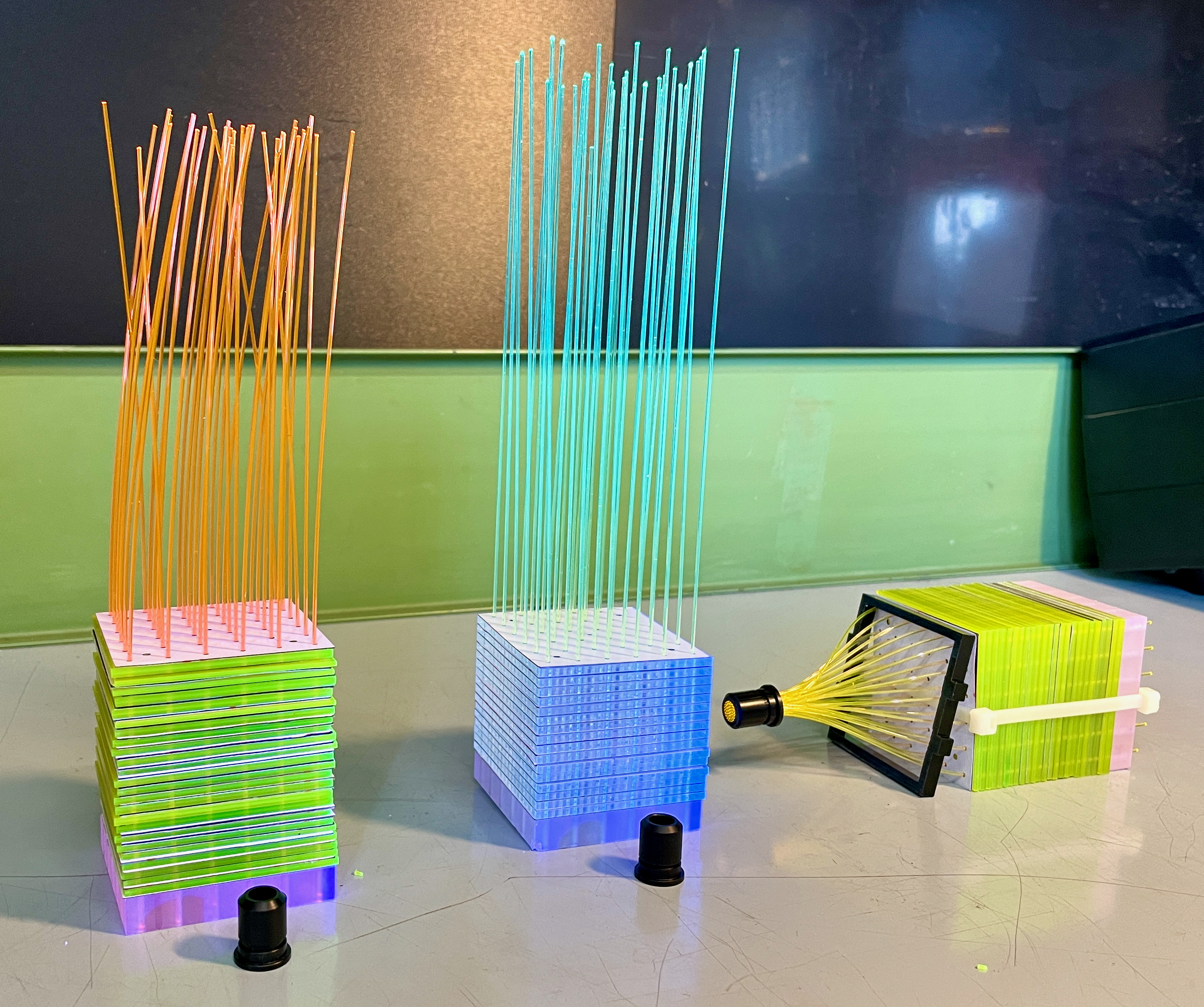}
\caption{Shashlyk modules for beam tests. The modules are $5.5\times5.5$~cm$^2$ in cross section and about 6~cm in depth. The module at the center uses a conventional blue scintillator and Y-11 fibers. The other modules use the NC scintillator and NCA-1 (left) and O-2 fibers (right).}\label{fig:shashlyk}
\end{figure}

On the basis of these results, we constructed and tested a small shashlyk test module with 0.2\% CsPbBr$_3$/polyacrylate nanocomposite read out with orange WLS fibers (Kuraray O-2, or a custom-dyed fiber produced at Kuraray, NCA-1), as well as an essentially identical module for comparison with conventional blue polystyrene scintillator from Protvino~\cite{Atoian} read out with green fibers (Kuraray Y-11), as shown in Fig.~\ref{fig:shashlyk}.
The tests were carried out at the CERN SPS in fall 2022. 
The light yield for single mips from the NC module was found to be only about 5\% of that from the conventional module.
Two possible hypotheses to explain the low light yield were identified: 1) excess self-absorption by the nanoparticles, and 2) inefficient excitation of the nanoparticles, whether because of their low concentration or because of the lack of an efficient channel for the transfer of energy to the nanoparticles from the polyacrylate matrix. 

To overcome the problem of self-absorption, we tested a new nanocomposite incorporating an additional stage of wavelength shifting. Instead of CsPbBr$_3$ nanocrystals, mixed-halide CsPb(Br,Cl)$_3$ nanocrystals (with about half of the Br atoms substitued with Cl atoms) were used. The resulting nanocrystals emit in the blue-violet, and coumarin-6 was used to shift the light back to the $\sim$520-nm green of the CsPbBr$_3$ nanocrystals for readout with the orange WLS fibers, reducing self-absorption. Unfortunately, the surface passivation of the nanocrystals destroyed during substitution reaction, leading to aggregation of the crystals in the nanocomposite, characterized by a milky appearance and poor transparency. A module constructed with this nanocomposite was tested together with the other modules at the CERN PS in spring 2023; unsurprisingly, the light yield was very low. The concept remains interesting, and especially with direct synthesis of the CsPb(Br,Cl)$_3$ nanocrystals to preserve the surface passivation, should be tried again.

\begin{figure}
\centering
\includegraphics[width=0.35\textwidth]{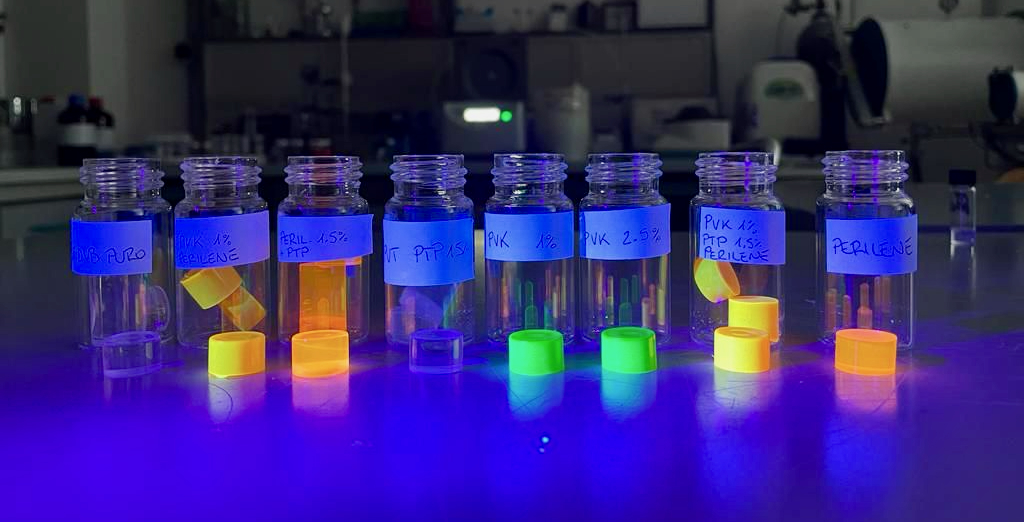}
\caption{Samples for the light yield measurements from the NC24 and Blank series, seen under ultraviolet light.}\label{fig:samples}
\end{figure}

After these initial tests of shashlyk modules, in order to simplify the evaluation of different NC formulations, we decided to work with smaller samples of scintillator, typically of about 1~cm$^3$ in volume, with direct light readout. In the meantime, we developed new protocols for the synthesis of thermally polymerized perovskite nanocomposites. This allows the use of an aromatic matrix material, e.g., polyvinyltoluene (PVT), as in conventional scintillators, providing a mechanism for energy transfer from the matrix to the nanocrystals via the primary scintillation light from the aromatic rings in the matrix. Table~\ref{tab:samples} lists 15 samples tested at the CERN PS with mips and at the Frascati BTF with 450-MeV electrons (which have mip-like energy deposit for small samples) in late 2023 and early 2024. Samples of both conventional molecular (including some commerical standards) and NC scintillator were tested for comparison, together with a number of control samples consisting of only the non-scintillating materials (the matrix materials, in some cases together with the WLS dyes used in some of the nanocomposites). Several of these samples are seen in Fig.~\ref{fig:samples}. All samples were standardized to the same dimensions (cylindrical, 14 mm diameter by 7 mm thick) and read out on one side with a Hamamatsu 13360-6050 SiPM. Tests with no samples present, positional scans, and reproducibility tests were also made. The analysis of the data collected in these tests is ongoing. We summarize our preliminary findings as follows. 

\begin{table*}[t]
\centering
{\footnotesize
\begin{tabular}{lll}
\hline\hline
Name & Composition & Notes \\
\hline
EJ200 & Proprietary & Commercial reference \\
EJ232Q & Proprietary & Commercial reference \\
Mol\_0 & 1.5\% PTP + 0.04\% POPOP & Blue, see \cite{Atoian} \\
Mol\_1 & 1.5\% PTP + 0.04\% BTP & Blue \\
NC23\_0 & 1.5\% PTP + 1.0\% CsPbBr$_3$ & Green \\
NC23\_1 & 1.5\% PTP + 1.0\% CsPbBr$_3$ + perylene & Orange \\
NC24\_0 & 1.0\% CsPbBr$_3$ & Green \\
NC24\_1 & 2.5\% CsPbBr$_3$ & Green \\
NC24\_2 & 1.5\% PTP + 1.0\% CsPbBr$_3$ & Green \\
NC24\_3 & 1.0\% CsPbBr$_3$ + perylene & Orange \\
NC24\_4 & 1.5\% PTP + 1.0\% CsPbBr$_3$ + perylene & Orange \\
Blank\_0 & Only matrix & Blank for NC24\_0 and NC24\_1 \\
Blank\_1 & 1.5\% PTP & Blank for NC23\_0 and NC24\_2 \\
Blank\_2 & perylene & Blank for NC24\_3 \\
Blank\_3 & 1.5\% PTP + perylene & Blank for NC23\_1 and NC24\_4\\
\hline\hline \\
\end{tabular}
}
\caption{Samples for light yield measurements. All samples except for the commercial samples from Eljen use 90\% PVT and 10\% divinylbenzene (DVB) as the matrix material. PTP: para-terphenyl, POPOP: 1,4-bis(5-phenyloxazol-2-yl)benzene, BTP: benzothiophene. ``Perylene'' refers to the perylene dyad described in \cite{Gandini}, used here as a green-to-orange WLS dye. The perovskite nanocrystals used for the two batches NC23 and NC24 were synthesized with different protocols and surface passivation techniques.}\label{tab:samples}
\end{table*}

Relative to EJ200, Blank\_1 (PVT + PTP) gives about 20\% of the light output, Blank\_0 (PVT only) gives about 7\% of the light output, and no sample at all gives about 3\% of the light output (arising from the direct response of the SiPM to the beam, possibly from Cherenkov photons from the entrance window or coupling grease). This provides a baseline for the activity of the components of the nanocomposite and test setup other than the perovskite nanocrystals themselves. Adding the perylene to either blank to shift the emission to orange does not change the light output from the blank. So, any absorption of light from the matrix or from the PTP by the perylene is compensated by the re-emission of light from the perlyene. The absorption of light by the perylene is presumably small, because if a large fraction of the primary light were absorbed/re-emitted by the perylene, we would see a small decrease in the overall response due to the decreased sensitivity of the SiPM at the longer wavelength.

Relative to EJ200, Mol\_0, a fairly standard formulation for a blue scintillator, with the same dyes used for the scintillator in the conventional shashlyk module (see~\cite{Atoian}), gives about 50\% of the light output, and EJ232Q about 33\%. 

Relative to Blank\_0, NC24\_0 and NC24\_1 each give about 64\% of the light output. Thus, independently of the concentration of CsPbBr$_3$ (1.0 vs 2.5\%), adding the perovskite lowers the light output with respect by the same factor, compared to the matrix alone.
Adding PTP increases the light output of PVT alone by 3.4x and of PVT with perovskite by 2.6--4.5x (with different results obtained for the samples from batches with different passivation/protocol). The PTP therefore plays an important role in transferring the deposited energy from the matrix to the perovskite.

Relative to Blank\_1, NC23\_0 gives about 86\% of the light output, while NC24\_2 (same composition, different passivation/protocol) gives only 49\% of the light output. Thus, for the samples with PTP, adding the perovskite lowers the light yield, by a larger factor for the second batch than for the first.
In general, the fact that adding perovskite decreases the light output suggests that the perovskite is absorbing its own light, in addition to the light from the PTP.
Adding perylene restores some of the light output in both cases (samples NC23\_1 and NC24\_4), increasing it by about 20\% relative to the samples without perylene (NC23\_0 and NC24\_2). This is consistent with the hypothesis that self-absorption by the perovskite plays a significant role in determining the LY: with perylene, at least some of the light is shifted to longer wavelengths before self-absorption.

For the NC23\_2 sample, some of the perovskite precipitated out on one face of the sample. When this sample is coupled with the precipitate layer in contact with the SiPM, the LY is lower than when the precipitate layer is not in contact with SiPM. In this case, when the beam is incident in a region with more precipitate, the response is lower. This is further evidence that perovskite self-absorption is blocking light output.

Spectral corrections were not performed for the purposes of these comparisons. The corrections are at maximum 20\% and making them does not change the above picture.

In summary, our results suggest that the use of perovskite quantum dots to make an NC scintillator for HEP is not straightforward. The realization of the promise of these materials for the creation of very fast, bright, and radiation-tolerant scintillators for HEP will require efforts to increase the light output, which is currently limited by the self-absorption of the perovskite nanocrystals. 

\section*{Acknowledgements}

This project has received funding from the European Union’s Horizon 2020 Research and Innovation program as a part of the AIDAinnova project, grant no. 101004761, the Horizon Europe EIC Pathfinder program as a part of the UNICORN project, grant no.~101098649, and the Italian Ministry of Universities and Research as part of the PRIN project IRONSIDE, project no.~2022RHCPFF, as well as from INFN CSN1 as R\&D for the NA62/HIKE experiment.

\bibliographystyle{elsarticle-num-names}
\bibliography{main}

\end{document}